# EnHMM: On the Use of Ensemble HMMs and Stack Traces to Predict the Reassignment of Bug Report Fields

Md Shariful Islam
Dept. Electrical and Computer Engineering
Concordia University
Montreal, QC, Canada
mdsha_i@ece.concordia.ca

Abdelwahab Hamou-Lhadj
Dept. Electrical and Computer Engineering
Concordia University
Montreal, QC, Canada
wahab.hamou-lhadj@concordia.ca

Korosh Koochekian Sabor
Dept. Electrical and Computer Engineering
Concordia University,
Montreal, QC, Canada
k_kooche@ece.concordia.ca

Mohammad Hamdaqa
Department of Computer and Software Engineering
Polytechnique Montreal
Montreal, QC, Canada
mhamdaqa@polymtl.ca

Haipeng Cai
School of Electrical Engineering and Computer Science
Washington State University
Pullman, WA, USA
haipeng.cai@wsu.edu

*Abstract*— Bug reports (BR) contain vital information that can help triaging teams prioritize and assign bugs to developers who will provide the fixes. However, studies have shown that BR fields often contain incorrect information that need to be reassigned, which delays the bug fixing process. There exist approaches for predicting whether a BR field should be reassigned or not. These studies use mainly BR descriptions and traditional machine learning algorithms (SVM, KNN, etc.). As such, they do not fully benefit from the sequential order of information in BR data, such as function call sequences in BR stack traces, which may be valuable for improving the prediction accuracy. In this paper, we propose a novel approach, called EnHMM, for predicting the reassignment of BR fields using ensemble Hidden Markov Models (HMMs), trained on stack traces. EnHMM leverages the natural ability of HMMs to represent sequential data to model the temporal order of function calls in BR stack traces. When applied to Eclipse and Gnome BR repositories, EnHMM achieves an average precision, recall, and F-measure of 54%, 76%, and 60% on Eclipse dataset and 41%, 69%, and 51% on Gnome dataset. We also found that EnHMM improves over the best single HMM by 36% for Eclipse and 76% for Gnome. Finally, when comparing EnHMM to Im.ML.KNN, a recent approach in the field, we found that the average F-measure score of EnHMM improves the average F-measure of Im.ML.KNN by 6.80% and improves the average recall of Im.ML.KNN by 36.09%. However, the average precision of EnHMM is lower than that of Im.ML.KNN (53.93% as opposed to 56.71%).

## I. Introduction

Bug reports (BRs) contain a wealth of information that is used by triaging and development teams to understand the causes of bugs in order to provide fixes. The problem is that, for various reasons, it is common to have BRs with missing or incorrect information, hindering the bug resolution process [1][19][21]. Xia et al. [30] showed that 80% of the BRs they analyzed (190,558 BRs in total) have their fields reassigned. Guo et al. [21] argued that the BR field reassignment problem is due to various factors including the difficulty to identify the root cause of a bug, ambiguous ownership of BR components, poor BR quality, difficulty to determine the proper fix, and workload balancing.

To address the BR field reassignment problem, researchers (e.g., [1][30][37]) have turned to machine learning techniques. The common practice is to build models that leverage historical BRs (the ground truth) to automatically predict when a given BR field should be reassigned. Existing approaches have mainly relied upon classifiers that make use of natural language in the title and description of the BRs. For example, Xia et al. [30] trained a multi-label imbalanced KNN model (Im-ML.KNN) that is based on BR field metadata, BR descriptions and summaries, and a combination of both. Although these techniques have shown to be useful, they fail to take advantage of the richly detailed sequential information that is present in stack traces included in BRs. A stack trace contains a sequence of function calls that are in memory when a bug occurs, which may be a better characterization of a bug as opposed to BR description, which is prone to errors related to the use of natural language.

In this paper, we propose an approach that uses sequences of function calls in stack traces and ensemble Hidden Markov Models (HMMs) to predict the reassignment of BR fields. HMM is a classification technique (more precisely a stochastic process) that is designed specifically to model sequential data [39]. HMMs are widely used in other areas such as intrusion detection [9][17][25], DNA processing [34], speech recognition [13], and image processing [40]. Our ensemble HMM approach, called EnHMM, combines multiple HMMs, trained by varying the number of hidden states, at the decision level. This design choice is inspired by prior studies in the field of anomaly detection (e.g., [17][27][28]), which showed evidence that the combination of multiple HMMs increases accuracy over a single HMM. We conjecture that a best-fit ensemble HMM model, trained on stack traces of reassigned and not reassigned BRs, would help predict the probability of an unknown BR field.

We applied EnHMM to BRs from the Eclipse and Gnome systems. For Eclipse, our approach provides an average precision, recall, and F-measure of 54%, 76%, and 60%, respectively. For Gnome, we obtained about 41% precision, 69% recall, and 51% F-measure. We also found that EnHMM improves over the best single HMM by 36% for Eclipse and 76% for Gnome. These results demonstrate that EnHMM, trained on BR stack traces, holds real promise for predicting BR field reassignments.

However, not all BRs come with stack traces. In our case study, only 12.9% and 19.08% of the studied Eclipse and Gnome BRs have stack traces. This is mainly due to the fact that many bug tracking systems are still not equipped with adequate mechanisms for managing traces. The objective of this study is not to replace the use of BR descriptions, but rather to demonstrate the viability of using information contained in stack traces to help improve predictive models for BR reassignment. We anticipate that future techniques will combine BR descriptions with trace information. This study should be seen as a step towards achieving this goal.

The remaining parts of the paper are as follows: In Section II, we provide background information on HMM and how to construct ensemble HMMs. In Section III, we present our approach to predict the reassignment of BR fields using stack traces and ensemble HMMs. In Section IV, we describe the experimental setup and results. In Section V, we discussed threats to validity, followed by related work in Sections VI, and finally, conclusion and future directions in Section VII.

## II. BACKGROUND

### A. Hidden Markov Models (HMM)

HMM is a statistical Markov model that is particularly useful for modeling sequential data (e.g., time series data). Fig. 1 illustrates a generic HMM model, $\lambda = (A, B, \pi)$, using an example sequence of function calls. The matrix A represents the state transition probability distribution, the matrix B represents the probability distribution of observation sequences, and $\pi$ represents the initial state probability distribution of each hidden state. Training an HMM using a discrete sequence of observations $\mathcal{O}\text{-}(\mathcal{O}_0, \mathcal{O}_1, \ldots, \mathcal{O}_{T-1})$ aims at maximizing the likelihood function $P(\mathcal{O}|\lambda)$ over the parameter space represented by $A, B$, and $\pi$ [17].

There exist several algorithms for learning the HMM parameters [12]. In our work, we use the Baum-Welch (BW) algorithm, which is the most commonly used Expectation-Maximization (EM) algorithm [3]. The BW algorithm iteratively uses a Forward-Backward (FB) algorithm [16] at each iteration to efficiently evaluate the likelihood function $P(\mathcal{O}|\lambda)$, and then updates the model parameters until the likelihood function stops improving or a maximum number of iterations is reached.

An HMM is a soft detector due to the fact that it gives a score (i.e., the likelihood probability) instead of a decision (in our case, a decision could be a specific BR field being reassigned or not). A soft detector can be converted into multiple crisp detectors (a detector in a binary classification problem that always gives a decision) by setting various thresholds on scores. We can then plot the performance of the crisp detectors, produced by one soft detector, into a Receiving Operating Characteristics (ROC) curve after setting various thresholds on scores. This way, a set of decisions produced by a crisp detector using a validation or a testing set is represented by a single point on the ROC curve where the x-axis represents the computed false positive rate (fpr) and the y-axis represents the true positive rate (tpr).

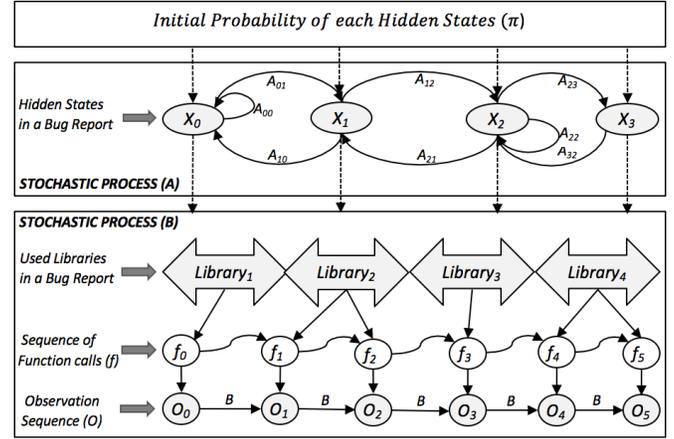

Fig. 1. A generic HMM model in training a BR field.

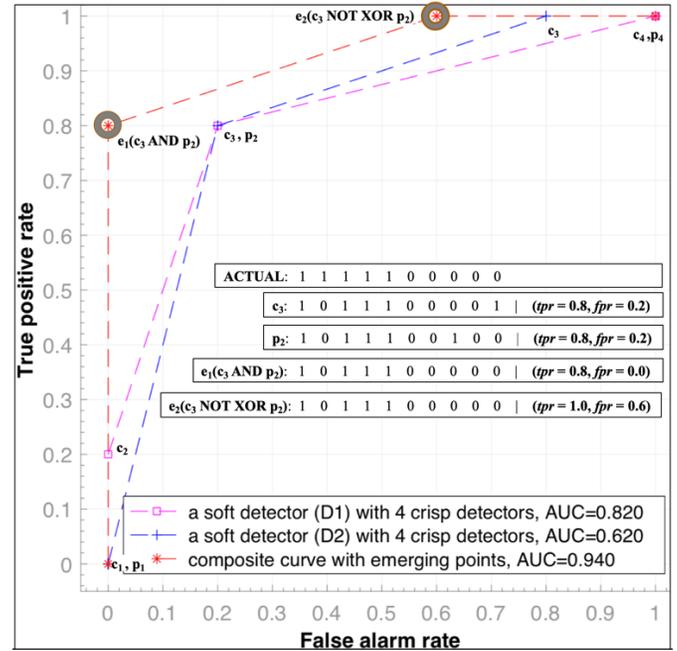

Fig. 2. An example of Boolean combination of HMMs.

Fig. 2 shows two soft HMM detectors, D1 and D2, which have four corresponding crisp detectors (i.e., single points on the ROC curve), obtained by setting four different thresholds on scores computed by D1 and D2. The two soft detectors, D1 and D2, produced two ROC curves where each has four candidate crisp detectors: D1(c1, c2, c3, and c4) and D2(p1, p2, p3, and p4). The Area Under the Curve (AUC) of the ROC curve produced by the soft detector D1 is 0.82, and D2 is 0.62, meaning that D1 performs better than D2. There is however a way to further improve accuracy by combining the decisions produced by both D1 and D2. This is the subject of the next subsection.

*B. Ensemble HMMs Using Boolean Combination*

Multiple HMMs can be generated by varying the number of hidden states. There exist various ways for combining classifiers (see [10]). In this paper, we focus on Boolean combination techniques. Barreno et al. [31] proposed an approach, called Pair-wise Brute-forces Boolean Combination (BBC2), for combining multiple detectors. BBC2 uses all possible combination pairs among all the available candidate crisp detectors. For example, in Fig. 2 the eight candidate crisp detectors (c1 to c4 and p1 to p4) produce 66 combination pairs. Each pair is then combined by ten different Boolean functions (a∧b, ¬a∧b, a∧¬b, ¬(a∧b), a∨b, ¬a∨b, a∨¬b, ¬(a∨b), a⊕b, a≡b). Therefore, it produces 66x10=660 emerging responses on the ROC space, which are then turned into 660 emerging points (e) on the ROC space. The points that have the highest AUC are then selected to compute the target composite ROC curve. In this example, two emerging points, e1 and e2 are used to compute the final composite ROC curve that improves the AUC.

Though effective, the BBC2 approach suffers from scalability problems due to the large number of required combinations. To address this issue, Khreich et al. [28] proposed the Iterative Boolean Combination (IBC) approach. IBC combines all available soft detectors in an iterative manner. In the first iteration, IBC starts by combining the first two soft detectors. The resulting composite ROC curve is then combined with the third soft detector, and continues up to combine the last soft detector. The process is repeated iteratively until no further improvement is obtained. At the end, IBC computes the final composite ROC curve and stores all the sequences of Boolean combination rules that are used to reach each of the emerging points (e) on that composite ROC curve. The combination rules are then used during testing.

Recently, Shariful et al. [17] proposed a new approach called Weighted Pruned Iterative Boolean Combination (WPIBC) that uses the Cohen's kappa statistics to define the level of (dis)agreement between two combined soft or crisp detectors [4][26]. The goal is to ensure the diversity among the combined detectors [14][17]. WPIBC prunes redundant detectors to reduce the computation time and to improve scalability, especially with a large number of combined detectors. We leverage WPIBC, in this paper, to combine multiple HMM models.

WPIBC works in three phases: Consider k HMMs ($HMM_1$, … $HMM_k$) soft detectors that produce $S_k$ set of scores on the validation set, V. Assume the possible threshold on scores $S_k$, produced by the soft $HMM_k$ is $T_k$. Phase 1 of WPIBC first selects $l$ most diverse soft HMM detectors out of k HMMs detectors using the weighted kappa agreement coefficient. Phase 2 selects $C_l$ complementary crisp detectors from $T_k$ possible candidate crisp detectors using simple kappa agreement coefficient. At the end, Phase 3 constructs the Boolean combination rules by combining those selected base soft and complementary crisp detectors using the same IBC Boolean combination technique. We also applied these three phases to construct an ensemble HMM from multiple HMM models. More details are provided in the next section.

### III. ENHMM APPROACH

Our approach for predicting the reassignment of BR fields consists of four phases as shown in Fig. 3: (1) preprocessing, (2) training, (3) validation, and (4) testing. In the preprocessing phase, we extract and profile sequences of function calls from stack traces of BRs. Note that not all BRs come with stack traces so we only include BRs with stack traces in our dataset. In the training phase, we use temporal sequences of function calls extracted from stack traces to train multiple HMMs for each BR field of interest (e.g., product, component, etc.). In the third phase, the validation phase, we select the most diverse detectors out of the available HMMs. For this, we use WPIBC [17], which ensures diversity among the combination of multiple detectors. The selected diverse detectors are used to construct the proposed ensemble HMMs. In the last phase, the testing phase, we use the constructed Boolean combination rules on each BR field of the testing set of BRs to predict whether it gets reassigned or not.

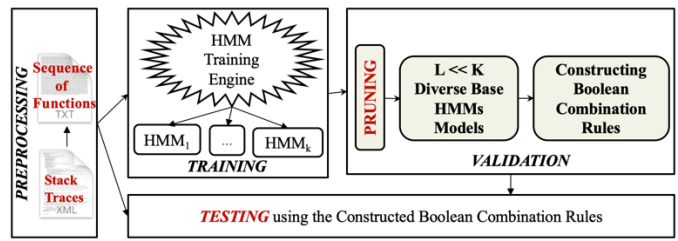

Fig. 3. An overview of our approach.

*A. Extracting and Profiling Sequences of Function Calls from Stack Traces*

A stack trace contains a sequence of function calls that are in memory when the crash occurs. In both Eclipse and Gnome bug tracking systems (used in this study), a BR submitter manually appends stack traces to BR descriptions and comments. To extract stack traces, we need to use regular expressions.

Bettenburg et al. [20] implemented a tool (Infozila) to extract stack traces from Eclipse BR descriptions and showed that their regular expression can extract stack traces with 98% accuracy. Lerch et al. [11] improved the regular expression proposed by Bettenburg et al. [20] to detect stack traces with a higher accuracy and proposed the following regular expression, which we use in our study:

[EXCEPTION] ([:][MESSAGE])? ([at][METHOD][([SOURCE] [)] )+ ( [Caused by:] [TEMPLATE] )?

Similarly, we need to define a regular expression to extract stack traces from BR descriptions in the Gnome bug tracking system. We designed the following regular expression after the third author examined manually over 100 Gnome BRs:

([#NUMBER] [HEX ADDRESS] [IN] [FUNCTION NAME] [(] [PARAMETERS] [)] ([FROM] | [AT]) ([LIBRARYNAME] | [FILENAME]))*

For each BR, we extract the sequence of function calls in its associated stack trace, which we will use to train multiple HMMs.

### B. Training an HMM

Our approach is used to predict the reassignment of any BR field of interest (e.g., component, product, severity, OS, version, etc.) that we refer to as BR field, $F_i$.

For a given $F_i$, we create an HMM by specifying the number of hidden states. The training phase consists of the following steps. We split the BRs into two sets: the BRs that have their field $F_i$ reassigned (R) and those that have their field $F_i$ not reassigned (NR). We use 70% of BRs from R to train the HMM. We use 10% of BRs from R and another 10% of BRs from NR to create the validation set. For testing (see the next subsection), we use 20% of BRs from R and the remaining 90% of BRs from NR. This way of splitting the data is a common practice in machine learning. This said, a different splitting may yield different results, which constitutes an internal threat to validity of our approach.

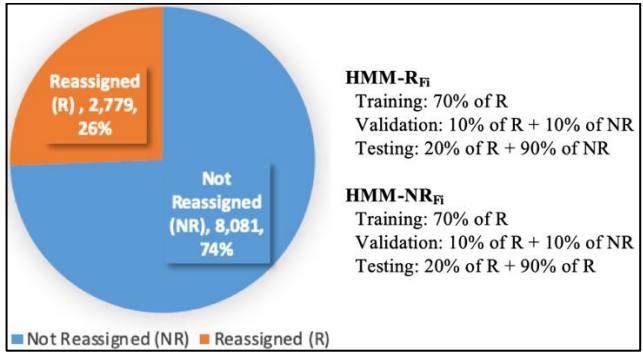

Fig. 4. Splitting the training, testing, and validation sets from the Eclipse bug reports on field, $F_i$ (i=Component) for HMM-$R_{Fi}$ and HMM-$NR_{Fi}$ models.

The output of this phase is an HMM that learns the pattern of BR-associated stack trace for which field $F_i$ is reassigned. We call this model HMM-$R_{Fi}$. This model can help predict for a new incoming BR whether field $F_i$ would get reassigned or not. However, the limited number of trained reassigned BRs (i.e., observations from the rare class) on a specific field $F_i$ causes a data imbalance problem as shown by Xia et al. [30]. Simply learning a model from the BRs for which Field $F_i$ is reassigned will most likely increase the false positive rate. To address this, we need to create another model that is trained on the major class observations (meaning BRs for which $F_i$ is not reassigned). We create another model, called HMM-$NR_{Fi}$ to represent BRs in the historical data for which $F_i$ is not reassigned. The idea is to combine multiple instances of each model by varying the number of hidden states (see next subsection) into a powerful detector that knows about both the rare and major class observations. HMM-$NR_{Fi}$ is trained using the same process as HMM-$R_{Fi}$. We use 70% of NR for training, 10% from R and another 10% from NR for validation. For testing, we use 90% of R BRs and 20% from NR. Fig. 4 shows how the data is split for training, validation, and testing purposes for both HMM-$R_{Fi}$ and HMM-$NR_{Fi}$ with an example of 10,860 BRs collected from the Eclipse project on 'component' field (given in Table I).

### C. Constrcuting Ensemble HMMs

The proposed ensemble HMMs are composed of HMM-$R_{Fi}$ and HMM-$NR_{Fi}$; each trained by varying the number of hidden states from N=10, 20…200. As a result, for each field $F_i$, we will have 20 HMM-$R_{Fi}$ and 20 HMM-$NR_{Fi}$ models combined. To our knowledge, there is no work that precisely defines how many hidden states we should have for best accuracy. Most studies (e.g., [28]) vary the number of hidden states as we propose in this paper.

The combination of these multiple HMM-$R_{Fi}$ and HMM-$NR_{Fi}$ soft detectors works at the decision label (i.e., '0' for not reassigned and '1' for reassigned). A decision is made by a crisp HMM-$R_{Fi}$ or HMM-$NR_{Fi}$ detector with a predefined threshold, $\theta$. Assume, in the validation set, we have $n$ BRs for Field $F_i$. We therefore obtain $n$ scores ($S_n$) computed by a trained soft HMM-$R_{Fi}$ / HMM-$NR_{Fi}$ detector. We obtain $n$ responses {$R_n$: 1 if $Sn > \theta$, otherwise 0}, which also represents the number of crisp detectors. Our HMM decision-level combination technique is based on WPIBC and consists of three steps (as shown in the Background section): (a) selecting base soft detectors, (b) selecting complementary crisp detectors, and (c) constructing Boolean combination rules.

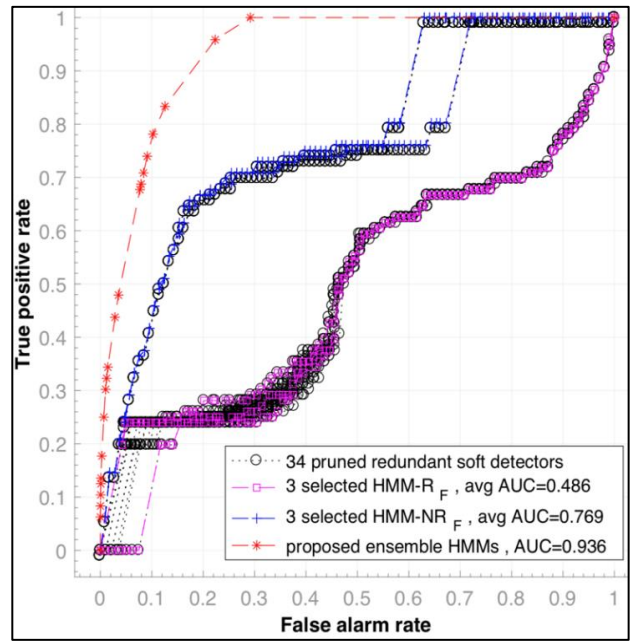

Fig. 5. Example of selected six diverse base HMM-$R_{Fi}$ and HMM-$NR_{Fi}$ soft detectors after pruning all the redundant ones under the ROC space using the validation set.

**Selecting Base Soft Detectors:** Suppose, there are k trained HMM-$R_{Fi}$ and HMM-$NR_{Fi}$ soft detectors and each one produces a set of scores ($S_k$) of size |V|, where V is the validation set. We use $T_k$ to refer to all possible thresholds on scores. Therefore, we have k ROC curves ($S_k, T_k$) with k AUC values. Initially, we select a *base* soft detector $k^* =$ max[AUC(k)] for which the AUC is the highest. Then we compute agreement coefficients between the base soft detector ($k^*$) and all the other soft detectors. We set an agreement threshold τ to 90% as a default value. This means that soft detectors that agree 90% with scores computed by the base soft detector ($k^*$) are considered redundant, and therefore should be pruned. Assume, we found k~ redundant copies of the base detector $k^*$. So, we select the base $k^*$ and prune k~ redundant ones. The process is repeated with the remaining ($k - k\text{\textasciitilde} - k^*$) soft detectors and continues until we are left with only one base soft detector. At the end, we obtain a total of l << k diverse base soft detectors.

TABLE I. STATISTICS ON BRS (BR) WITH STACK TRACES COLLECTED FROM ECLIPSE AND GNOME BUG REPOSITORIES.

| Dataset | Class Label | Assignee | | Product | | Component | | Version | | OS | | Priority | | Severity | | Status | |
|---|---|---|---|---|---|---|---|---|---|---|---|---|---|---|---|---|---|
| | | #BR | % | #BR | % | #BR | % | #BR | % | #BR | % | #BR | % | #BR | % | #BR | % |
| Eclipse | Not-Reassigned | 3,566 | 33 | 9,156 | 84 | 8,081 | 74 | 8,875 | 82 | 10,194 | 94 | 9,702 | 89 | 9,593 | 88 | 9,451 | 87 |
| | Reassigned | 7,294 | 67 | 1,704 | 16 | 2,779 | 26 | 1,985 | 18 | 666 | 6 | 1,158 | 11 | 1,267 | 12 | 1,409 | 13 |
| Gnome | Not-Reassigned | 3,752 | 73 | 8,813 | 83 | 7,930 | 75 | 6,612 | 63 | 10,471 | 99 | 9,404 | 89 | 9,317 | 88 | 9,736 | 92 |
| | Reassigned | 6,827 | 27 | 1,766 | 17 | 2,649 | 25 | 3,967 | 37 | 108 | 1 | 1,175 | 11 | 1,262 | 12 | 843 | 8 |

Fig. 5 shows an example with k=40 trained soft detectors (i.e., 20 HMM-$R_{Fi}$ and 20 HMM-$NR_{Fi}$) using the validation set. We can see that only six (i.e., l=6, three from HMM-$R_{Fi}$ and three from HMM-$NR_{Fi}$) soft detectors are selected as diverse. All the other ones are pruned because they are redundant. The resulting l=6 base soft detectors are then used to select the final complementary crisp detectors.

**Selecting Complementary Crisp Detectors:** Suppose we have $T_l$ possible thresholds on scores computed by a base soft $HMM - R_{Fi}^l$ or $HMM - NR_{Fi}^l$ detector (l). We therefore obtain $T_l$ candidate crisp $HMM - R_{Fi}^l(T_l)$ or $HMM - NR_{Fi}^l(T_l)$ detectors. Then, we compute kappa (kp) agreement coefficients between each crisp detector's decisions and decisions from the ground truth. The accurate crisp detectors should be close to $kp \approx kp_{max}$ and their complementary crisp detectors should be close to $kp \approx kp_{min}$. Assume the number of selected crisp detectors is D and the ratio between accurate and their complementary crisp detectors is 50%, we sort candidate crisp detectors in a descending order based on their kp agreement coefficients. Then, we select the top D/2 (i.e., 50% of total) as accurate crisp detectors and the bottom D/2 as their complementary ones, respectively.

**Constructing Boolean Combination Rules:** We combine decisions/responses (0/1) produced by each selected complementary crisp detector by leveraging the WPIBC Boolean combination technique [17]. WPIBC uses the same Boolean operators as previous approaches, namely IBC [28], except that it uses only base soft detectors with their selected complementary crisp detectors instead of all available candidate soft and crisp detectors (as it is the case of IBC). We also use ten different Boolean combination functions to combine two crisp detectors' decisions on the ROC space. Initially, we combine the first two base soft detectors and then, the resulting emerging responses are combined with the next base one and so on. We repeat this combination process iteratively until no further improvement is reached. The composite ROC curve (red curve in Fig. 5) with the AUC about 93% is the combination of selected complementary crisp HMM-$R_{Fi}$/HMM-$NR_{Fi}$ detectors produced by six selected base soft HMM-$R_{Fi}$/HMM-$NR_{Fi}$ detectors using the validation set and θ as a threshold. The constructed Boolean combination rules are then used during testing.

## IV. CASE STUDY SETUP AND RESULTS

This case study aims to answer the following questions:

- RQ1: How does EnHMM perform in terms of its ability to predict BR field reassignment?
- RQ2: How does EnHMM perform in comparison to a single HMM when predicting BR field reassignment?
- RQ3: How does EnHMM compare to Im.ML.KNN, a leading technique?

### A. Datasets

We use Eclipse and Gnome bug repositories to assess the performance of our approach. Eclipse and Gnome are two open source software systems and their bug repositories are publicly available through Bugzilla bug tracking system. We only consider BRs with status "resolved", "closed", "verified", and "fixed". For Eclipse, we collected 83,984 BRs that were submitted between January 2008 to July 2011, which is the same period that was used in other studies (e.g., Im-ML.KNN [30], ML.KNN [18]). The number of Eclipse BRs with stack traces is 10,860, which accounts for 12.9% of the total number of BRs. For Gnome, we collected 55,438 BRs from December 2007 to July 2011, among which 10,579 (19.08%) have stack traces. This dataset was used by the authors in other studies. (We are currently building larger datasets on which we intend to replicate this work.)

Table I shows the distribution of reassigned and not reassigned BRs for eight BR fields: Product, Component, Version, OS, Priority, Severity, and Status. As expected, the number of BRs for which field $F_i$ is not reassigned is much higher than the number of BRs that are reassigned, which shows a clear imbalance of the data. As we explained in Section III.B, we address this by creating a model for each class, HMM-$R_{Fi}$ and HMM-$NR_{Fi}$, and combine them.

### B. Training HMMs for Field Fi

As discussed in Section III, to train an HMM, we split the BRs associated with field ($F_i$) into two groups: BRs that have $F_i$ reassigned, and those that have $F_i$ not reassigned. Each group is then divided into three sets: training (70%), validation (10%), and testing (20%). The 10% validation set contains BR traces from each group. For testing, we use 20% of BR traces from the training class and 90% from the other group of BR traces. For example, in Eclipse, the number of stack traces used for training, validation, and testing one HMM-$NR_{Fproduct}$ model, given that the number of BRs with stack traces that have the product field reassigned and not reassigned is 1,704 and 9,156, respectively (see Table I) is as follows:

- Training set contains 6,409 traces (=9,156*70%)
- Validation set contains 1,086 traces (9,156*10% + 1,704*10%)
- Testing set contains 3,365 traces (=9,156*20% + 1,704*90%)

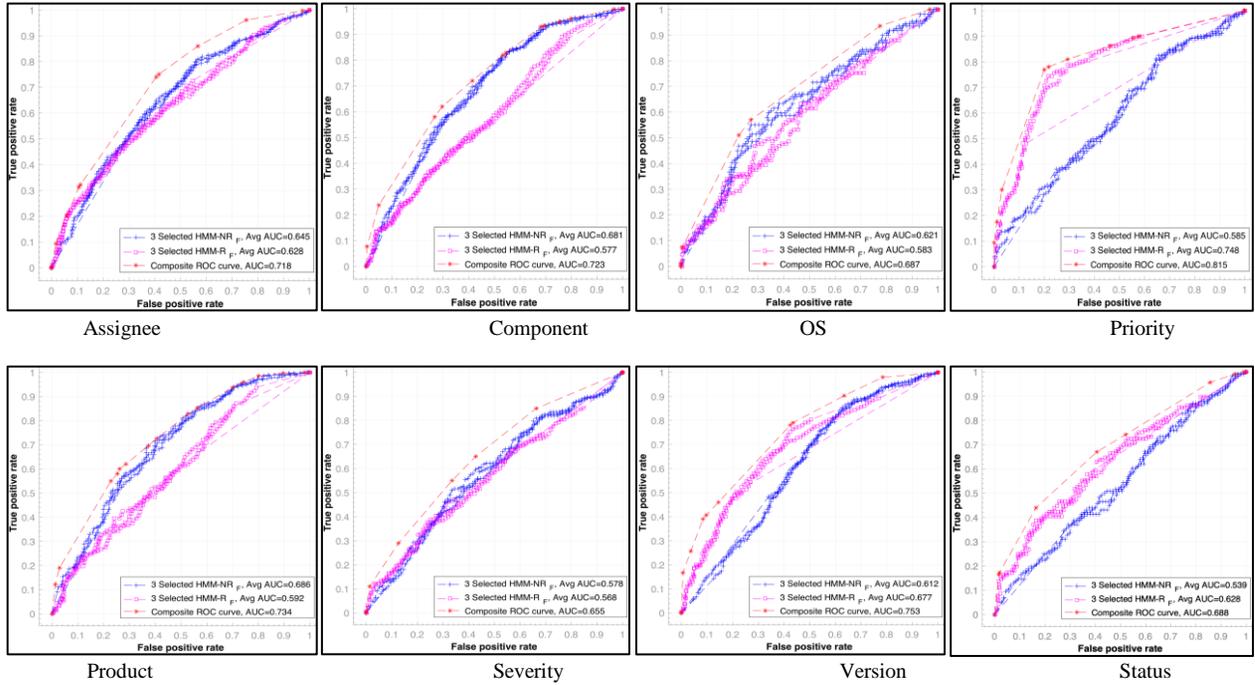

Fig. 6 Results on the testing set for Eclipse bug report fields

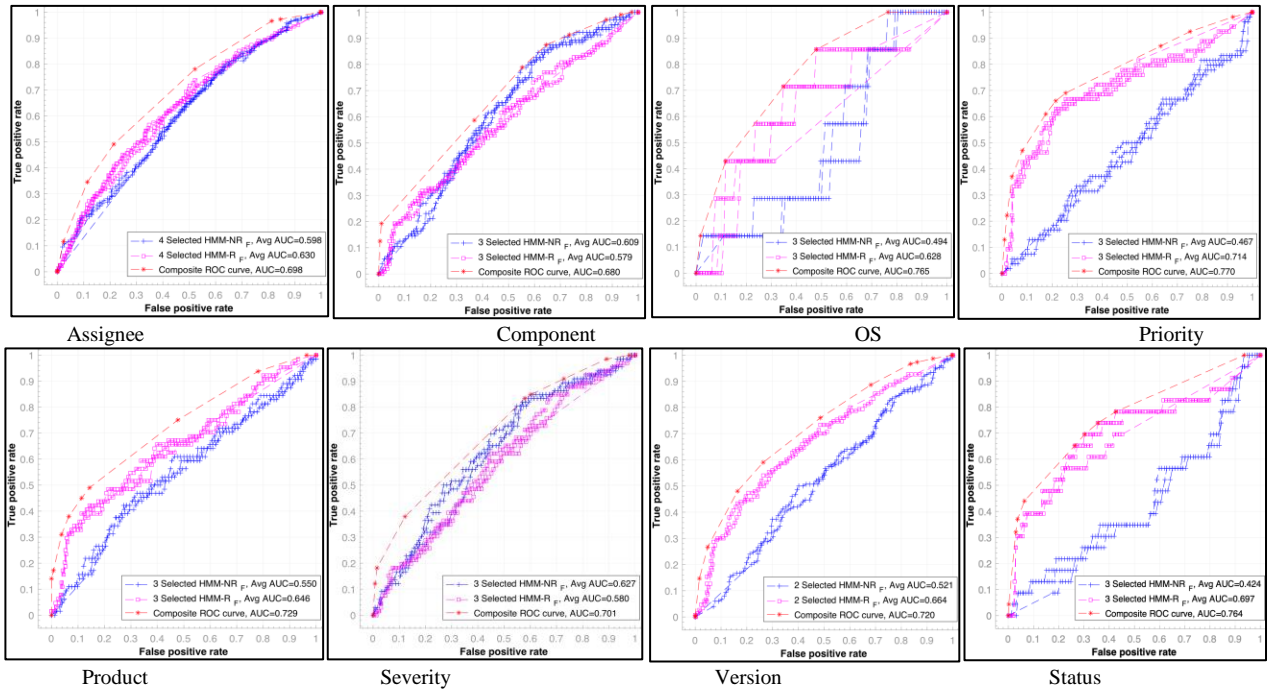

Fig. 7 Results on the testing set for Gnome bug report fields.

We apply the same process to HMM-R$_{Fproduct}$ and also to construct HMM-R$_{Fi}$ and HMM-NR$_{Fi}$ for every other field $F_i$. In addition, for each field $F_i$, we train 20 HMM-R$_{Fi}$ and HMM-NR$_{Fi}$ by varying the number of hidden states (N), from 10 to

200 with bonds of 10. In total, we built 280 (=40*7) different HMM models for the prediction of the seven BR fields shown in Table I. Note that not all of these HMM models are used in the actual prediction since the WPIBC (the selected HMM combination approach) prunes the redundant ones.

*C. Evaluation Metrics*

In addition to the ROC curve that we discussed in Section III, we also use precision, recall, and F-measure to measure the performance of EnHMM to predict BR field reassignment. These metrics are used in the literature to evaluate the accuracy of a classifier [1][8][18][21].

Precision and recall are defined as follows:

$$Precision = \frac{TP}{TP + FP} \quad (1)$$

$$Recall = \frac{TP}{TP + FN} \quad (2)$$

TP: True Positives; FP: False Positives; FN: False Negatives.

Precision is the ratio of the number of BRs that we correctly predicted that their field (Fi) is reassigned (TP) to the total number of BRs for which we predicted that their field (Fi) is reassigned (TP+FP). Recall is the ratio of the number of BRs that we correctly predicted that their field (Fi) is reassigned (TP) to the total number of BRs that actually have their field (Fi) reassigned (TP+FN). To have a better perception of the result, we also use F-measure, a harmonic mean of precision and recall and is defined as follows:

$$F - measure = \frac{2 \times Precision \times Recall}{Precision + Recall} \quad (3)$$

*D. Experimental Results*

We use the ROC curves (see Fig. 6 and Fig. 7) to show the effectiveness of EnHMM in predicting whether a BR field of a new incoming BR would be reassigned or not by addressing RQ1, RQ2, and RQ3.

TABLE II. ACCURACY OF ENHMM

| BR Field | Datasets | Precision | Recall | F-measure |
|---|---|---|---|---|
| Assignee | Eclipse | 80.15% | 97.12% | 87.82% |
|  | Gnome | 82.69% | 95.91% | 88.82% |
| Component | Eclipse | 62.50% | 67.87% | 65.00% |
|  | Gnome | 45.61% | 100.0% | 62.65% |
| OS | Eclipse | 36.82% | 100.0% | 53.83% |
|  | Gnome | 28.71% | 100.0% | 55.81% |
| Priority | Eclipse | 54.75% | 75.63% | 63.52% |
|  | Gnome | 26.32% | 55.56% | 35.71% |
| Product | Eclipse | 57.57% | 98.90% | 72.78% |
|  | Gnome | 45.61% | 40.63% | 42.98% |
| Severity | Eclipse | 21.04% | 72.87% | 32.66% |
|  | Gnome | 22.17% | 65.15% | 33.08% |
| Version | Eclipse | 61.19% | 72.00% | 66.16% |
|  | Gnome | 50.88% | 58.00% | 54.21% |
| Status | Eclipse | 57.41% | 26.72% | 36.47% |
|  | Gnome | 28.57% | 34.78% | 31.37% |
| **Average** | **Eclipse** | **53.93%** | **76.39%** | **59.78%** |
|  | **Gnome** | **41.32%** | **68.76%** | **50.59%** |

**RQ1. How does EnHMM perform in terms of its ability to predict BR field reassignment?**

We can easily compute the best precision, recall, and F-measure for each predicted BR field $F_i$ from the corresponding ROC curve shown in Fig. 6 and Fig. 7. Each point (*fpr*, *tpr*) on the final composite ROC curve produced by EnHMM represents the predicted responses (i.e., the decisions whether the testing BRs will be reassigned (i.e., 1) on field $F_i$ or not reassigned (i.e., 0) on field $F_i$. We used this set of predicted responses (i.e., a set of points) on the composite ROC curve for Field $F_i$ to compute a set of precisions, recalls, and F-measures using Equations (1), (2), and (3). Finally, a point (i.e., the *tpr* and *fpr* of the responses or predicted outcomes) out of all the points on the ROC curve produced by EnHMM (red one with star marker points) that give the maximum F-measure is selected as the best predictor with a best precision, recall, and F-measure for each BR field $F_i$.

Table II shows the best F-measure of the proposed ensemble HMMs for each field $F_i$ from the corresponding ROC curve shown in Fig. 6 for Eclipse and Gnome datasets. Overall, EnHMM performs relatively well for most cases, with some noticeable exceptions. For example, it only detects the "severity" field with a precision of 21% for Eclipse and 22% for Gnome (the lowest precision obtained). We also notice that for the "status" field, EnHMM achieves a low recall for both Eclipse and Gnome (27% and 35% respectively). This may be due to the low number of BRs for which this field is reassigned as shown in Table I. On the other hand, we notice a very high precision and recall for fields that contain a large number of BRs for which the respective field is reassigned very often. For example, the "assignee" field, which is reassigned in 68% of the BRs for Eclipse and 27% BRs in Gnome can be predicted with 80% precision and 97% recall for Eclipse and 83% precision and 96% recall for Gnome. We need to conduct more studies to understand the reasons behind the performance of EnHMM by examining various factors including the impact of the size of the dataset on the approach, as well as the size and content of the BR stack traces. For now, we state the following finding:

> **Finding 1:**
>
> EnHMM achieves an average precision, recall, and F-measure of 54%, 76%, and 60% on Eclipse dataset and 41%, 69%, and 51% on Gnome dataset.

**RQ2. How does EnHMM perform in comparison to a single HMM when predicting BR field reassignment?**

From Fig. 6 and Fig. 7, we can see that EnHMM (represented with the red curve in the figures) always gives a better accuracy than the best selected single HMM detector (the blue/pink curves) for all BR fields for both datasets. The ensemble HMMs significantly improves the AUC, while reducing the false positive rates compared to the best single HMM (the ROC curve in blue or pink depending on the field, which is the closest to the EnHMM red curve). For example, for the "assignee" field in Eclipse data (see Fig. 6 Assignee), the AUC of the ROC curve corresponding to the three selected HMM-NR$_{assignee}$ is 0.645, the AUC of the three selected HMM-R$_{assignee}$ is 0.628, and the AUC of EnHMM (composite ROC curve) = 0.718. This also shows that the rules constructed by the ten different Boolean combination functions yields good results.

To dig deeper, we analyzed each ROC curve shown in Fig. 6 and Fig. 7 on Eclipse and Gnome testing datasets to find the maximum *tpr* at the y-axis against a maximum tolerable *fpr (MTPR)* at the x-axis for each BR field using EnHMM and a single HMM. We measure the improvement as follows:

Improvement = (TPR$_{EnHMM}$ – TPR$_{singleHMM}$) / TPR$_{singleHMM}$

Table III shows the results. For example, for the "assignee" field in Eclipse data, the maximum tolerable FPR (MTFPR) is 12%, the TPR obtained using EnHMM that corresponds to MTFPR in the ROC curve is 32% and that of a single HMM is 26%, which shows that EnHMM results in 23% (i.e., (32%-26%)/26%) improvement over the best single HMM.

TABLE III. IMPROVEMENT OF ENHMM OVER A SINGLE HMM

| BR Field | Datasets | MTFPR | TPR EnHMM | TPR Single HMM | Improvement |
|---|---|---|---|---|---|
| Assignee | Eclipse | 12% | 32% | 26% | 23% |
|  | Gnome | 11% | 34% | 27% | 26% |
| Component | Eclipse | 5% | 24% | 14% | 71% |
|  | Gnome | 1% | 19% | 4% | 375% |
| OS | Eclipse | 22% | 51% | 47% | 9% |
|  | Gnome | 12% | 43% | 43% | 0% |
| Priority | Eclipse | 2% | 30% | 22% | 36% |
|  | Gnome | 8% | 47% | 42% | 12% |
| Product | Eclipse | 2% | 19% | 12% | 58% |
|  | Gnome | 14% | 49% | 42% | 17% |
| Severity | Eclipse | 12% | 29% | 20% | 45% |
|  | Gnome | 12% | 38% | 20% | 90% |
| Version | Eclipse | 10% | 41% | 30% | 37% |
|  | Gnome | 5% | 26% | 15% | 73% |
| Status | Eclipse | 16% | 44% | 39% | 13% |
|  | Gnome | 6% | 44% | 39% | 13% |
| **Average** | **Eclipse** | **10%** | **34%** | **26%** | **36%** |
|  | **Gnome** | **9%** | **38%** | **29%** | **76%** |

In addition, Fig. 6 and Fig. 7 show the number of selected detectors out of the 40 detectors (20 HMM-R$_{Fi}$ and 20 HMM-NR$_{Fi}$) used initially for each field. For example, for the "product", "component", "severity" and "assignee" fields in Eclipse dataset, our approach only needed 6 detectors (3 HMM-R$_{Fi}$ and 3 HMM-NR$_{Fi}$) out of 40 to provide optimum AUC (=0.734). The maximum number of selected detectors (i.e., after the pruning step) independently from any field is 8. We needed a maximum of 5 HMM-R$_{Fi}$ and 3 HMM-NR$_{Fi}$ to attain best accuracy for the prediction of the OS and Priority fields. Similarly, we needed 3 HMM-R$_{Fi}$ and 3 HMM-NR$_{Fi}$ to predict the "component", "OS", "product", "priority" and "severity" fields for the Gnome dataset. In other words, our approach only needed a maximum of 8 out 40 initial detectors (20%) to predict any of the fields, which suggests that it is not only effective for predicting the reassignment of these fields, but also scalable with the growing number of detectors.

**Finding 2:**

EnHMM improves over a single HMM by 36% for Eclipse and 76% for Gnome. In addition, EnHMM requires at most 20% of the initial detectors thanks to the Kappa-based pruning approach used to prune redundant detectors.

## RQ3: How does EnHMM compare to existing techniques?

We compare our approach to a recent approach proposed by Xia et al. [29], called the imbalanced multi-label k-Nearest Neighbors (Im-ML.KNN). The authors proposed a machine learning method that is a composite classifier where each classifier uses the same multi-label KNN (ML.KNN) machine learning algorithm [18] to train the model. The main novelty of Im-ML.KNN is the combination of three classifiers that are built on top of three separate features types: BR field metadata, BR description and summary, and a mix of both. When applied to four large BRs datasets (OpenOffice, Netbeans, Eclipse, and Mozilla) containing a total of 190,558 BRs, the authors showed that their approach achieves an average F-measure score of 56%-62%. They also showed that Im-ML.KNN improves on average the F-measure scores by 119.69%, 9.11%, and 161.08% when compared with past methods namely the method proposed by Lamkanfi et al. [1], ML.KNN [30], and HOMER-NB [8], respectively.

The authors, however, did not provide a reproduction package, which made it challenging for us to reuse their approach. Reimplementing Im-ML.KNN would require resources and even if we succeeded to do so, it would have been difficult to reproduce their experiments on our datasets because of the number of parameters that we needed to provide, which we could not find (at least explicitly) in the corresponding papers. In addition, the only common dataset between their approach and ours is the Eclipse dataset.

Despite these challenges, we attempt, in this paper, to provide a baseline comparison by comparing the results of our approach when applied to the Eclipse BRs with stack traces (this represents only 12.9% of BRs of the entire Eclipse dataset) to the results obtained by Im-ML.KNN applied to the entire Eclipse dataset as reported in their respective papers. We want to note that in their paper [29], the authors reported that they collected 50,639 BRs for almost the same period (i.e., Jan 2008 to July 2011), whereas we found that during this period there are 83,984 BRs with status "fixed". This discrepancy may be due to the fact that we included BRs with status "verified". We do not think that this has an impact on the comparison since we are using BRs with stack traces, which is a small subset of the entire BR space anyway.

Table IV shows the best F-measures of EnHMM for each BR field and that of Im.ML.KNN. We also measure the improvement. As we can see that, although EnHMM is tested on far fewer data points than Im.ML.KNN, the average F-measure score of EnHMM improves the average F-measure score of Im.ML.KNN by 6.80% (this is calculated as follows: (59.78%-55.97)/55.97%).

TABLE IV. COMPARISON BETWEEN ENHMM AND IM.ML.KNN BASED ON F-MEASURE

| F-measure | Average | Assignee | Component | OS | Priority | Product | Severity | Version | Status |
|---|---|---|---|---|---|---|---|---|---|
| EnHMM | 59.78% | 87.82% | 65.00% | 53.83% | 63.52% | 72.78% | 32.66% | 66.16% | 36.47% |
| Im-ML.KNN | 55.97% | 86.67% | 63.65% | 66.06% | 54.13% | 73.34% | 25.77% | 63.41% | 14.75% |
| **Improvement** | **6.80%** | **1.33%** | **2.12%** | **-18.51%** | **17.35%** | **-0.76%** | **26.74%** | **4.34%** | **147.25%** |

The EnHMM F-measure score is higher than that of Im.ML.KNN for five fields out of eight. Major improvements are observed for the "priority", "severity", and "status" fields (between 17.35% to 147.25%). Slight improvements can be seen for the "assignee", "component", and "version" fields (between 1.33% and 4.34%). For the "OS" field, we observe that EnHMM F-measure score is considerably lower than that of Im.ML.KNN (improvement of -18.51%), possibly because of the low number of reassigned BRs used for training (only 6% as shown in Table I). This also suggests that having more BRs with stack traces may improve the accuracy of the proposed solution. We intend to conduct more studies to understand the underlying reasons behind the performance of EnHMM across these BR fields. We need to examine in more depth how the size of the dataset, the quality of the traces, and the use of a particular learning algorithm impact the results.

Table V shows a comparison of both approaches using the average precision and recall. Xia et al. [29] did not report the precision and recall obtained by applying Im.ML.KNN to each field. They only included the averages shown in Table V. We can see that, on average, EnHMM has a much higher recall (76.39% compared to 56.13%), but a lower precision (53.93% compared to 56.71%).

TABLE V. COMPARISON BETWEEN ENHM AND IM.ML.KNN

| Approach | Average Precision | Average Recall |
|---|---|---|
| EnHMM | 53.93% | 76.39% |
| Im-ML.KNN | 56.71% | 56.13% |
| **Improvement** | **-4.90%** | **36.09%** |

**Finding 3:**

The average F-measure score of EnHMM, trained on 12.9% of Eclipse BRs, improves the average F-measure of Im.ML.KNN when trained on the entire dataset described in [29] by 6.80%. EnHMM improves the average recall of Im.ML.KNN by 36.09%. The average precision of EnHMM is lower than that of Im.ML.KNN by an improvement of -4.90%.

*E. Discussion*

**On the performance of EnHMM:** The appealing results obtained by EnHMM are attributable to the power of HMMs in modeling sequential data as opposed to traditional machine learning techniques, which do not take full advantage of sequential data. Moreover, fusing weak and best classifiers using 10 different Boolean functions maximizes the diversity between two combined detectors. In fact, it is the most important ground truth for any ensemble approach [14][17].

**On the use of stack traces:** Our findings clearly demonstrate the viability of the use of stack traces in predicting bug report fields. This confirms the need to better collect, store, and manage stack traces whenever a bug report is submitted. For the present time, both Eclipse and Gnome rely on stack traces that are copied and pasted in BR descriptions by end users. This process is error-prone and may result in the presence of noise. Bug report tracking systems must be equipped with powerful mechanisms for managing historical traces that can later be used for all types of applications including the prediction of BR field reassignment.

**Precision vs Recall:** As we mentioned earlier, EnHMM improves recall significantly but does not necessarily improve precision. In other words, with EnHMM, we can predict BR fields better than Im.ML.KNN, with a risk of having higher rate of false positives (number of correct BR fields that are predicted as possible reassignments). A high positive rate may not be desirable since users may lose trust in the system when they see many false alarms. The low precision may be due to many factors including the small size of BRs with stack traces, the fact that EnHMM relies solely on stack traces unlike Im.ML.KNN, which combines BR field metadata, BR description and summary, and a mix of both. We should also consider in practice to implement a feedback loop that can help our prediction algorithm to learn new cases so as to prevent from misclassifying newer and similar cases. Finally, we can also work on improving the algorithmic part of EnHMM by combining heterogenous classifiers.

**On the use of heterogenous detectors**: EnHMM is based on a combination of multiple HMM homogenous classifiers, trained by varying the number of hidden states. This said, the combination process itself is not linked to the sole use of HMM. It can, for example, be used to combine decisions from other types of classifiers such as those built using SVM, KNN, etc. as discussed by Khreich et al. [38]] in their anomaly detection approach. We believe that this can further improve the diversity aspect of the combination process (which is now supported through the use of the Kappa coefficient).

V. THREATS TO VALIDITY

**Threats to external validity:** Our approach is evaluated against two open source datasets. We need to experiment with more datasets that contain a large number of stack traces to generalize the results. We also need to use other features such as BR descriptions, summaries, and so on to assess the effectiveness of EnHMM on these features in situations where one cannot rely on stack traces. In addition, the comparison section is based on two different sets of BRs from Eclipse bug reports that were submitted between Jan 2008 and July 2011. It is provided here as a baseline comparison to position our approach with respect to the literature. A fair comparison must be based on the exact datasets.

**Threats to internal validity:** In our approach, the way we set the hyperparameters A and B, conditional probability matrices, to construct HMM could be a threat to internal validity. We used the validation set to optimize A and B. A different validation set could result in a different initialization of A and B, which my produce another model. However, to our knowledge there is no clear solution to this problem and most studies that use HMM follow random initialization of A and B and repeat this process several times until a satisfactory model is obtained. In addition, we chose to build 40 HMMs for each BR field by varying the number of hidden states. A different configuration may yield other results. Another threat may be with respect to the use of regular expressions to extract stack traces from BR descriptions. Our regular expression may have missed some stack traces, which may impact the accuracy of our approach. In addition, we implemented many scripts to extract data, build HMMs, etc. Although care was exercised when writing these scripts, errors may have occurred.

**Threats to construct validity:** The construct validity shows how the used evaluation measures could reflect the performance of our predictive model. In this study, we used precision, recall, F-measure, ROC curves, and AUC. These measures are widely used in similar studies to assess the accuracy of machine learning models.

## VI. RELATED WORK

There exist many studies that mine BRs for various purposes (e.g., [7][22]). The closest work to our study is that of Xia et al. [29]. The authors built a model to predict reassignment of BR fields using multi-label learning algorithm (ML.KNN). Their method (Im-ML.KNN) combines three different classifiers based on BR field metadata, BR descriptions and summaries, and a combination of these features. Their approach achieved an accuracy (F-measure) ranging from 56% to 62%. Bettenburg et al. [19] conducted a survey among developers and users of Apache, Eclipse, and Mozilla to understand what makes a good BR. They showed that since users are not primarily technical domain experts, they cannot choose BR fields correctly. They found that the steps to reproduce and stack traces are the most useful fields in BRs. Incomplete information in BRs appears to be one of the problems encountered by developers to fix the bugs.

Guo et al. [21] showed that there are five main reasons for BR field reassignment: Finding the root cause, determining ownership, identifying the root cause (proper fix determination), poor BR quality (incorrect or incomplete BRs), and workload balance. They showed that imprecise BR fields lead to the BR being transferred between development teams. They referred to this fact as the bug pong concept. They also showed that the incorrect selection of BR fields, increases the bug fixing time. Breu et al. [23] showed that BR questions can be categorized into eight groups: Missing information, clarification of information provided, information for triaging, information needed for debugging, information on how to provide corrections, status inquiry, resolution, and administration questions. They also showed that incorrect information is the main cause of triaging uncertainties.

Shihab et al. [5][6] showed that BRs that are reassigned take in average two times longer to be fixed. Sureka [2] showed that the Assignee field is the most reassigned field in the bug repositories. He applied a probabilistic model to the title and description of BRs to predict faulty component fields. The approach could be used to predict faulty component field of BRs with 42% accuracy. Lamkanfi et al. [1] showed that faulty component field of Eclipse and Mozilla BRs are frequently reassigned. They trained a Naïve Base classifier to predict reassignment of the component field of BRs in Eclipse and Mozilla based on BR component, reporter, operating system, version, severity, and summary. They showed that their approach achieves an accuracy of 44% for predicting if a bug will be reassigned and 83% if a bug will not be reassigned.

Several studies focused on using stack traces to detect duplicate BRs [11][33]. These studies build feature vectors based on the functions in stack traces. They showed that predictive models built based on stack traces can detect duplicate BRs with an accuracy of up to 90%. Other studies focused on using stack traces to predict BR fields including BR severity. Sabor et al. [32][35][36][37] built feature vectors based on the functions in stack traces. They showed that traces and BR categorical feature provide good accuracy.

## VII. CONCLUSION

In this paper, we proposed an effective approach for predicting the reassignment of BR fields to help improve the bug fixing process, and hence contributing to alleviate the costly burden of software maintenance activities [15]. Our approach, EnHMM, combines multiple HMMs using WPIBC, an anomaly detection algorithm that uses Boolean combination of classifiers, pruned using the Kappa coefficient. When applied to the Eclipse and Gnome BR repositories, EnHMM achieves an average precision, recall, and F-measure of 54%, 76%, and 56% on Eclipse dataset and 41%, 69%, and 51% on Gnome dataset. A preliminary comparison study shows that EnHMM achieves on average a better recall than im.ML.KNN, a leading BR field reassignment prediction method, but a lower precision. We can enhance precision in various ways: (a) increase the size of the training set by having more BRs with stack traces, (b) add other features such as BR field metadata and/or BR descriptions and summaries (if deemed of good quality), (c) implement a feedback loop to prevent misclassifying newer and similar cases, and (d) combining other types of classifiers such as SVM, KNN, etc. In the future, we aim to investigate how other sequential learning methods such as Long Short-Term Memory networks (LSTM) can be applied. We also aim to investigate other ensemble methods beyond those based on Boolean combination (see [10] [24]) and assess their impact on predicting BR field reassignments. In addition, the improvement obtained by EnHMM varies from one field to another. We need to dig deeper to understand what are the most important factors that affect accuracy for each field. Finally, an important future work is to apply EnHMM to larger datasets including datasets from the industry.